\documentclass[prl,twocolumn,nofootinbib,preprintnumbers,amssymb,amsfonts,amsmath,superscriptaddress,showpacs,hyperref]{revtex4-1}

\usepackage{graphicx}
\usepackage{bm}
\usepackage{color}
\usepackage{url}
\usepackage{epstopdf}

\newcommand{\keV}{{\rm keV}}

\newcommand{\GeV}{{\rm GeV}}

\newcommand{\muG}{\mu{\rm G}}

\newcommand{\Mpl}{M_{\rm Pl}}

\begin{document}

\title{Scale hierarchy in Ho{\v r}ava--Lifshitz gravity: a strong constraint from synchrotron radiation in the Crab nebula}

\author{Stefano Liberati}
\affiliation{SISSA, Via Bonomea 265, 34136, Trieste, Italy {\rm and} INFN, Sezione di Trieste}

\author{Luca Maccione}
\affiliation{Arnold Sommerfeld Center, Ludwig-Maximilians-Universit\"at, Theresienstra{\ss}e 37, 80333 M\"unchen, Germany} 
\affiliation{Max-Planck-Institut f\"ur Physik (Werner-Heisenberg-Institut), F\"ohringer Ring 6, 80805 M\"unchen, Germany} 

\author{Thomas P.~Sotiriou}
\affiliation{SISSA, Via Bonomea 265, 34136, Trieste, Italy {\rm and} INFN, Sezione di Trieste}

\begin{abstract}
Ho{\v r}ava--Lifshitz gravity models contain higher order operators suppressed by a characteristic scale, which is required to be parametrically smaller than the Planck scale. We show that recomputed synchrotron radiation constraints from the Crab nebula suffice to exclude the possibility that this scale is of the same order of magnitude as the Lorentz breaking scale in the matter sector. This highlights the need for a mechanism that suppresses the percolation of Lorentz violation in the matter sector and is effective for higher order operators as well.
\end{abstract}

\preprint{LMU-ASC 45/12, MPP-2012-109}
\maketitle


It has often been suggested  that Lorentz symmetry might be violated in quantum gravity 
(see {\em e.g.} \cite{Liberati:2012tb} for a review). 
Recently, the perhaps bolder proposal that abandoning Lorentz symmetry might be the payoff for rendering gravity renormalizable \cite{arXiv:0901.3775} has received much attention. The underlying idea is to modify the graviton propagator in the ultraviolet (UV) by adding to the action terms containing higher order spatial derivatives of the metric, but  refrain from adding higher order time derivatives in order to preserve unitarity.  

This procedure can take place straightforwardly once space-time has been foliated into a family of spacelike surfaces, labeled by the $t$ coordinate and with $x^i$ being the coordinate on each surface. However, once this preferred foliation has been imposed one cannot require invariance under general coordinate transformations. The resulting theory, called Ho\v rava--Lifshitz (HL) gravity (see {\em e.g.~}\cite{Sotiriou:2010wn} for a brief review), can still be invariant under the reduced set of coordinate transformations, or more rigorously diffeomorphisms, that leave this foliation intact, $t\to \tilde{t}(t)$ and $x^{i}\to \tilde{x}^{i}(t,x^i)$. Power counting renormalizability requires that the action includes terms with at least 6 spatial derivatives in 4 dimensions \cite{arXiv:0901.3775,Sotiriou:2010wn,arXiv:0902.0590,arXiv:0912.4757}. All lower order operators compatible with the symmetry of the theory are expected to be generated by radiative corrections, so the most general action takes the form \cite{arXiv:0909.3525}
\begin{equation}
\label{SBPSHfull}
S_{HL}= \frac{M_{\rm Pl}^{2}}{2}\int dt d^3x \, N\sqrt{h}\left(L_2+\frac{1}{M_\star^2}L_4+\frac{1}{M_\star^4}L_6\right)\,,
\end{equation}
where $M_{\rm Pl}$ denotes the Planck scale, $h$ is the determinant of the induced metric $h_{ij}$ on the spacelike hypersurfaces,
\begin{equation}
L_2=K_{ij}K^{ij} - \lambda K^2 
+ \xi {}^{(3)}\!R + \eta a_ia^i\,,
\end{equation}
where $K$ is the trace of the extrinsic curvature $K_{ij}$, ${}^{(3)}\!R$ is the Ricci scalar of $h_{ij}$, $N$ is the lapse function, and $a_i=\partial_i \ln N$. 
$L_4$ and $L_6$ denote a collection of 4th and 6th order operators respectively and $M_\star$ is the scale that suppresses these operators. It is perhaps tempting to call $M_\star$ the Lorentz breaking scale, but the theory exhibits Lorentz violations (LV) at all scales, as $L_2$ already contains LV operators. These Infrared (IR) Lorentz violations are controlled by three dimensionless parameters that take the values $\lambda=1$, $\xi=1$ and $\eta=0$ in General Relativity (GR).

Due to the reduced symmetry with respect to GR, the theory propagates an extra scalar mode. If one chooses to restore diffeomorphism invariance, then this mode manifests as a foliation-defining scalar field \cite{Jacobson:2010mx,Sotiriou:2011dr}.

From an IR perspective this theory can be viable and consistent for suitable choices of the dimensionless parameters $\lambda$, $\xi$ and $\eta$ \cite{arXiv:1007.3503}. An unappealing feature, however, is that $L_4$ and $L_6$ contain a very large number ($\sim 10^2$) of operators and independent coupling parameters. In remedy of this situation, restrictions to the theory have been proposed, which would limit the proliferation of independent couplings. 

One such restriction, leading to ``projectable HL gravity'' (see \cite{Weinfurtner:2010hz,Mukohyama:2010xz} for reviews),  is to impose that $N=N(t)$. Even though this version of HL gravity has only 9 independent couplings \cite{Sotiriou:2009gy}, the scalar mode is plagued by instabilities and exhibits strong coupling at unacceptably low energies \cite{Charmousis:2009tc,Sotiriou:2009bx,Blas:2009yd,Koyama:2009hc}. This signals the need to treat at least part of the theory non-perturbatively \cite{Mukohyama:2010xz,Gumrukcuoglu:2011ef}, which, however, jeopardizes the perturbative renormalizability arguments that served as the initial motivation for HL gravity. For this reason we do not consider the projectable version here.
Other restrictions can be imposed on the action, which do not alter the field content of the theory, see {\em e.g.}~\cite{Vernieri:2011aa}. However, we prefer to leave $L_4$ and $L_6$ unspecified here, as our analysis does not hinge on their exact form. 

An important scale in the theory is  $M_\star$. Obviously, $M_\star$ is bound from below from observational constraints which allow a minimum suppression scale $M_{\rm obs}$. Much more surprisingly, it is also bound from above \cite{Papazoglou:2009fj}. This is because the IR part of the action, $L_2$, exhibits strong coupling for the scalar mode at a scale $M_{\rm sc}$, which is parametrically smaller than $M_{\rm Pl}$, and whose size is controlled by the parameters $\lambda$, $\xi$ and $\eta$ \cite{Papazoglou:2009fj, arXiv:1003.5666}. Avoiding strong coupling (in order to not compromise perturbative renormalizability) requires $M_{\rm sc} > M_\star$ \cite{arXiv:0912.0550}. 

Taking into account the constraints on $\lambda$, $\xi$ and $\eta$ one has $M_{\rm sc} < 10^{16}$ GeV,  unless the parameters of the theory are tuned to satisfy certain bonds \cite{arXiv:1007.3503} , and the overall picture can be summarized as
$M_{\rm obs} <M_\star < 10^{16}\,\, {\rm GeV}$.
The exact value of $M_{\rm obs}$ depends strongly on what observations one intends to use. Gravity-related observations could lead to $M_{\rm obs}\sim$ few meV (sub mm tests), which would leave a very large window open for $M_\star$. 

However, even if one assumes that LV is confined to the gravitational sector at tree level, it is well known \cite{Collins:2004bp,Iengo:2009ix} that radiative corrections to the matter fields' propagators will induce LV terms into the matter dispersion relations at some scale $M_{\rm LV}(M_{\star})$. 
There are then 2 options: (a) $M_\star=M_{\rm LV}$, i.e.~$M_\star$ is a universal scale; (b) $M_\star\ll M_{\rm LV}$.  (Current phenomenological constraints already rule out the case $M_{\rm LV} \ll M_{\star}$ \cite{Liberati:2012tb,Jacobson:2005bg,Liberati:2009pf}.) Clearly, option (b) requires some mechanism which suppresses the percolation of LVs in the matter sector. Such mechanisms have been discussed in \cite{Pospelov:2010mp}.  Here, we will highlight the necessity for such mechanisms by focussing on option (a) and demonstrating that, in this case, matter LV constraints lead to $M_{\rm obs}\gtrsim M_{\rm sc}$, thus closing the available window for $M_\star$. 

Note that one can argue that some protective mechanism is needed already to shield  matter lower order operators (mass dimension three and four) from percolation of higher order LV, see {\emph e.g.}~\cite{Liberati:2012tb,Jacobson:2005bg,Liberati:2009pf,Pospelov:2010mp,GrootNibbelink:2004za,Bolokhov:2005cj}. However, such arguments depend on renormalization computations and eventual assumptions about the LV in the matter sector at the tree level. We shall opt here for a more direct argument that is based on higher order operators. This would also demonstrate the necessity for a mechanism that is not limited to lower order operators. 

{\em Framework.---} 
For what comes next we shall then assume that lower order operators in the matter sector are indeed protected. 
We shall also assume that the CPT and Parity (P) invariance of the gravitational action is preserved in the matter sector. Indeed if no CPT and P odd operators are present in the matter sector at the tree level one would not expect them to be generated via radiative corrections induced by the gravitational, CPT and P even, terms. This assumption forbids  helicity dependent terms and allows only even power of the momentum in the matter dispersion relation. 
These properties imply the equality of the matter and antimatter dispersion relations in this framework. So in the end, we will take the dispersion relations in the matter sector to be of the form
 \begin{equation}
E^{2} = m^{2} + p^{2} + \eta \frac{p^{4}} {M_{\rm LV}^{2}} +O\left(\frac{p^{6}}{M_{\rm LV}^{4}} \right)\, .
\label{eq:disp-rel}
\end{equation}

Missing extra suppression factors, one would naively expect the dimensionless coefficient $\eta$  to be of $O(1)$ at the Lorentz breaking scale and to be mildly different from such value at different energies due to the logarithmic renormalization group flow~\cite{Bolokhov:2005cj}. 

Dispersion relations of this order have been studied in the past (see e.g.~\cite{Jacobson:2005bg,Liberati:2009pf,Liberati:2012tb} for review) and several constraints were cast. Due to the relatively high order of the LV breaking terms, significant deviations from standard physics should be expected only at sufficiently high energies. In a threshold reaction such energy would normally be $p\approx  (m^2M_{\rm LV}^2/\eta)^{1/4}$ corresponding to 70 PeV for electrons and 3 EeV for protons for $M_{\rm LV} = \Mpl$ and $\eta=1$. 

Such high energies are indeed accessible thanks to Ultra High Energy Cosmic Rays (UHECR) physics and constraints on $\eta$ of $O(10^{-8})$ and $O(10^{-6})$ were deduced respectively on electron/photons and protons, assuming that the so called GZK cutoff~\cite{gzk} was detected in the spectrum of the UHECR~\cite{Liberati:2009pf,Liberati:2012tb}. Note that these numerical values were derived assuming $M_{\rm LV}=M_{\rm Pl}$ and they then imply $M_{\rm LV}\gg M_{\rm Pl}$ if one insists in having $\eta\approx O(1)$ in \eqref{eq:disp-rel}. Hence, the above mentioned option (a) $M_\star=M_{\rm LV}$ would be observationally non viable.

However, both these constraints rely on the crucial assumption that UHECRs were mostly constituted by protons. Nowadays, the interpretation of observational results on the development of UHECR showers in the atmosphere seems to favor a somewhat heavier composition of UHECRs, weakening the evidence for the detection of a pure GZK feature, thereby invalidating previous constraints.  It is then justified to not rule out option (a) solely on the base of this observational evidence.  As a result, we are not in position to cast any other effective constraints on the QED sector with $p^4$ corrections to the dispersion relation and we are left with very weakened ones of the hadronic sector. 
  A limit $M_{\rm LV}\gtrsim M_{\rm Pl}$ could still be placed by considering LV effects in the propagation of UHECR heavy nuclei \cite{Saveliev:2011vw}, however this would be based on a few simplifying assumptions about the percolation of LV from the fundamental level to nuclear physics.

 Setting aside UHECR physics, one might look back at the best constraints cast so far on  mass dimension 5 operators and see if they can be somewhat extended to the case of  mass dimension 6 operators, taking into account  that under assumption (a)  $M_{\rm LV}$ appearing in Eq.~(\ref{eq:disp-rel}) will have to be parametrically smaller than $M_{\rm Pl}$. Following this lead we shall then focus on one of the most efficient and robust constraints on dimension five operators in LV QED, i.e.~the one provided by the synchrotron radiation emitted by the most reliably known pulsar wind nebula: the Crab Nebula (CN).

{\em LV Synchrotron radiation.---} It was firstly realized in \cite{Jacobson:2002ye} that the synchrotron emission by ultra-relativistic electrons can be severely affected by LV due to the extreme sensitivity of its group velocity. In what follows we shall generalize the standard treatment to a dispersion relation of the kind \eqref{eq:disp-rel} of arbitrary order $n$
 \begin{equation}
E^{2} = m^{2} + p^{2} + \eta \frac{p^{n}} {M_{\rm LV}^{n-2}}\, .
\label{eq:disp-rel-n}
\end{equation}
As usual we shall neglect the Lorentz breaking term in the photon as such term is always largely negligible with respect to the electron one at the relative energies involved (up to 0.1 GeV for the photon against energies of the order of $10^3$ TeV for the electron).

It can be shown \cite{Jacobson:2002ye} that the typical synchrotron critical frequency $\omega_{c}$ in the LV case can be computed as %
\begin{equation}
\omega_{c}(E) = \frac{3}{2}eB\frac{\gamma^{3}(E)}{E}
\end{equation}
Assuming Hamiltonian dynamics, the group velocity of an electron with a modified dispersion relation \eqref{eq:disp-rel-n} can be computed as
\begin{equation}
v(E) = \frac{\partial E}{\partial p} \simeq 1-\frac{m^{2}}{2p^{2}} + \frac{n-1}{2}\eta\left(\frac{p}{M_{\rm LV}}\right)^{n-2}
\end{equation}
neglecting higher order terms.  We immediately see that $v(E)$ can exceed $1$ if $\eta > 0$ or it can be strictly less than $1$ if $\eta < 0$ once the LV term is prevailing on the suppressed mass term. This introduces a fundamental difference between particles with positive or negative LV coefficient $\eta$ and the possibility to see the effect even for $p/M_{\rm LV}\ll 1$.

{\em Constraints.---}
In \cite{Jacobson:2002ye} a simple constraint was cast using the fact that subluminal dispersion relations ($\eta<0$) admit a maximal critical frequency for the synchrotron emission. For arbitrary $n>2$ this would be
\begin{equation}
\omega_{c}^{max,(n)}=\frac{3eB}{2m}\left(1-\frac{4}{3n}\right)^{3/2}\!\!\left(\frac{4\, (M_{\rm LV}/m)^{n-2}}{-\eta(n-1)(3n-4)}\right)^{2/n}.
\end{equation}
A constraint was also obtained for $n=3$ by requiring this maximal critical frequency to be larger than the maximal observed frequency in the CN synchrotron spectrum, $\omega_{\rm obs}\approx 0.1$ GeV. We can run the same argument here for $n=4$ and straightforwardly derive a constraint $\eta \gtrsim -10^{5}$ (assuming $B\sim300~\muG$ and $M_{\rm LV} = \Mpl$), which would correspond to $M_{\rm LV} > 3\times10^{16}~\GeV$. 

While promising, this is not a double sided constraint. In order to be sensitive to the full range of the $\eta$ parameters and take into account competing LV and LI effects a much deeper analysis is needed. This was performed in \cite{Maccione:2007yc} where the possible LV induced modifications to the standard Fermi mechanism (which is thought to be responsible for the formation of the spectrum of energetic electrons in the CN) were considered and the synchrotron spectrum of the CN was recomputed taking into account all the new, LV induced, phenomena (such as vacuum \v{C}erenkov and helicity decay) for $n=3$ LV. 

\begin{figure}[tbp]
\begin{center}
\includegraphics[width=0.5\textwidth]{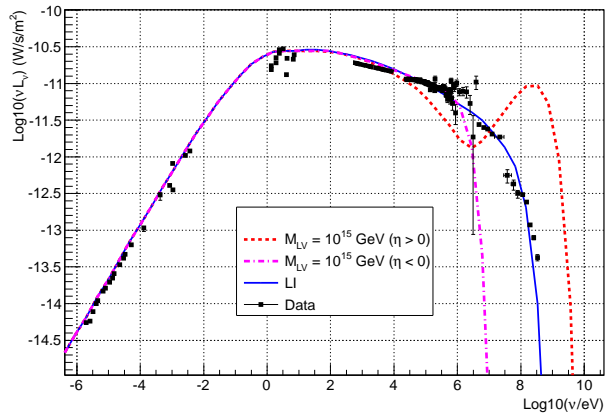}
\caption{Crab Nebula spectrum for the LI case (blue, solid curve), for the LV case $n=4$, with $M_{\rm LV} = 10^{15}~\GeV$ and $\eta>0$ (red, dashed curve), and for the case with same parameters but $\eta<0$ (magenta, dot-dashed curve). While, as discussed, the $\eta<0$ case would lead to premature fall off of the synchrotron spectrum, we see here that for $\eta>0$ there is a sudden surge of emission at high frequencies, followed by a dramatic drop due to the onset of vacuum \v{C}erenkov emission at the characteristic threshold energy $E_{\rm th}\approx \sqrt{mM_{\rm LV}}/\eta^{1/4}$.}
\label{fig:spectrum}
\end{center}
\end{figure}

We rework and extend here the numerical algorithm used in \cite{Maccione:2007yc} so to compute the broad-band spectrum of the Crab Nebula below 100 MeV down to radio frequencies for the modified dispersion relation \eqref{eq:disp-rel}. 
As in \cite{Maccione:2007yc} we fix the free parameters of the model (electron/positron density and spectrum and magnetic field strength) so that we reproduce the low energy part of the spectrum, which is not affected by LV (see Fig.~\ref{fig:spectrum}). We then consider how LV affects the higher energy part of the spectrum $(E\gtrsim100~\keV)$ and use $\chi^{2}$ statistics to measure when deviations from the observed spectrum due to LV become unacceptably large. 

\begin{figure}[tbp]
\begin{center}
\includegraphics[width=0.5\textwidth]{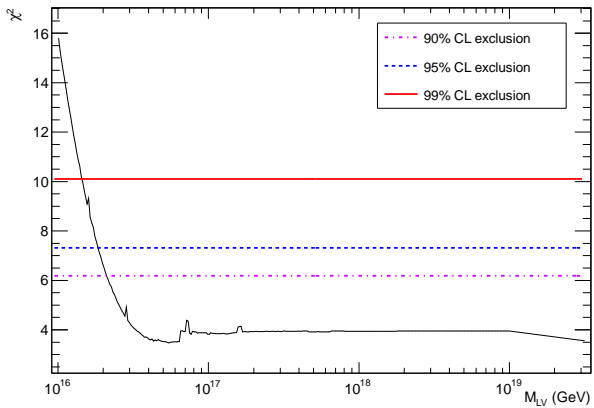}
\caption{Plot of the evolution of the reduced $\chi^{2}$ with the mass scale $M_{\rm LV} $. Horizontal lines show exclusion levels at 90\% (violet, dot-dashed), 95\% (blu, dashed) and 99\% (red, solid) CL. As already found in a previous analysis \cite{Maccione:2007yc}, we find that LV can improve the fit of the Crab Nebula spectrum with respect to a pure LI model. This is however due to both the LI model not reproducing all the spectral features properly, and the LV model containing the additional free parameter $\eta$.}
\label{fig:exclusion}
\end{center}
\end{figure}

As a paradigmatic example of our results we show in Fig.~\ref{fig:spectrum} a comparison of the LI spectrum with the LV spectra for $M_{\rm LV} = 10^{15}~\GeV$ and both $\eta>0$ and $\eta<0$.
We then show in Fig.~\ref{fig:exclusion} the dependence of the reduced $\chi^{2}$ on $M_{\rm LV}$.  By considering the offset from the minimum of the reduced $\chi^{2}$ we set exclusion limits at 90\%, 95\% and 99\% Confidence Level (CL), according to \cite{pdg}. Mass scales $M_{\rm LV} \lesssim 2\times10^{16}~\GeV$ are excluded at 95\% CL.

{\em Conclusions.---} In this Letter 
we placed constraints on LV in the matter sector in HL models by exploiting the broad band spectrum of the Crab Nebula. We obtain $M_{\rm LV}\gtrsim 2\times 10^{16}~\GeV$, assuming CPT and P invariance to be preserved in the matter sector. Would P invariance be broken, we would need to consider a more complicated problem involving two possibly different LV mass scales $M_{\rm LV 1,2}$, in a similar way as done in \cite{Maccione:2007yc}. Remarkably, the overall constraints would be of the same order of magnitude as the one we found here, although some part of the parameter space with $M_{\rm LV 1} \ll M_{\rm LV 2} \gtrsim 10^{16}~\GeV$ would still be allowed.

Lacking a detailed model for the percolation of LV from the gravity to the matter sector, the constraint derived above cannot be directly transferred to a limit on the characteristic scale of HL gravity $M_{\star}$. However, given that generically $M_{\star}<10^{16}~\GeV$ in order to avoid strong coupling, our current constraints appear incompatible with the possibility that  $M_{\rm LV}\sim M_{\star}$. Therefore a mechanism, suppressing the percolation of LV in the matter sector, must be present in HL models, and such mechanism should not only protect lower order operators. 

 As mentioned earlier, and has been discussed in \cite{arXiv:1007.3503}, if the parameters $\lambda$, $\xi$ and $\eta$ are tuned to satisfy certain bonds,  $M_{\rm sc}$ can become significantly higher.  If a justification for such a choice where to be found, then this would be a possible way out. Alternatively, one could resort to breaking of P invariance and allowing for a strong hierarchy between the two scales $M_{\rm LV1}$ and $M_{\rm LV2}$.

A much more appealing  option is offered by the mechanism proposed in \cite{Pospelov:2010mp}, which one might call of ``gravitational confinement". This mechanism assumes that no LV is present at the tree level in the matter sector (so to avoid the need for additional custodial symmetries like supersymmetry~\cite{GrootNibbelink:2004za,Bolokhov:2005cj}) and that $M_{\star}\ll M_{\rm Pl}$. In this case radiative corrections will percolate LV operators from the gravity sector to the matter ones but the gravitational coupling $G_N\sim M^{-2}_{\rm Pl}$ will do so by introducing strong suppression factors of the order $(M_{\star}/M_{\rm Pl})^{2}$. In \cite{Pospelov:2010mp} it has been shown that dimension 4 operators of matter can be efficiently screened from LV this way in HL models and this is expected to be the case for higher order operators as well. Our work motivates a detailed study of the efficiency of this mechanism.

{\em Acknowledgments:} We thank D.~Mattingly, M.~Pospelov, Y.~Shang and S.~Sibiryakov for useful comments on an earlier version of this manuscript. 
LM acknowledges support from the AvH foundation. TPS acknowledges partial financial
support provided under a Marie Curie Career Integration
Grant and the ``Young SISSA Scientists Research
Project'' scheme 2011-2012.

\end{document}